\documentclass{emulateapj}
\newcommand{\ms}{m s$^{-1}$ }

\usepackage{graphicx}
\usepackage{color}
\usepackage{amsmath}
\usepackage{amssymb}
\usepackage{hyperref}
\usepackage{epstopdf}
\usepackage{upgreek}
\usepackage{xcolor}
\hypersetup{
    colorlinks,
    linkcolor={red!80!black},
    citecolor={blue!80!black},
    urlcolor={blue!80!black}
}
\shorttitle{\lq{}Modal-noise\rq{} in single-mode fibers: a cautionary note}
\shortauthors{Halverson et al.}
\slugcomment{Accepted in ApJ Letters}

\begin{document}

\title{\lq{}Modal-noise\rq{} in single-mode fibers: A cautionary note for high precision radial velocity instruments}

\author{
Samuel Halverson\altaffilmark{1,2,3}, 
Arpita Roy\altaffilmark{1,2,3},
Suvrath Mahadevan\altaffilmark{1,2,3}, 
Christian Schwab\altaffilmark{4,5}
}

\altaffiltext{1}{Department of Astronomy \& Astrophysics, The Pennsylvania State University, 525 Davey Lab, University Park, PA 16802, USA; {shalverson@psu.edu}}
\altaffiltext{2}{Center for Exoplanets \& Habitable Worlds, University Park, PA 16802, USA}
\altaffiltext{3}{Penn State Astrobiology Research Center, University Park, PA 16802, USA}
\altaffiltext{4}{Macquarie University, Sydney, NSW 2109, Australia}
\altaffiltext{5}{Australian Astronomical Observatory, Sydney, Australia}


\begin{abstract}
Exploring the use of single-mode fibers (SMFs) in high precision Doppler spectrometers has become increasingly attractive since the advent of diffraction-limited adaptive optics systems on large-aperture telescopes. Spectrometers fed with these fibers can be made significantly smaller than typical \lq{}seeing-limited\rq{} instruments, greatly reducing cost and overall complexity. Importantly, classical mode interference and speckle issues associated with multi-mode fibers, also known as \lq{}modal noise\rq{}, are mitigated when using SMFs, which also provide perfect radial and azimuthal image scrambling. However, these fibers do support multiple polarization modes, an issue that is generally ignored for larger-core fibers given the large number of propagation modes. Since diffraction gratings used in most high resolution astronomical instruments have dispersive properties that are sensitive to incident polarization changes, any birefringence variations in the fiber can cause variations in the efficiency profile, degrading illumination stability. Here we present a cautionary note outlining how the polarization properties of SMFs can affect the radial velocity measurement precision of high resolution spectrographs. This work is immediately relevant to the rapidly expanding field of diffraction-limited, extreme precision RV spectrographs that are currently being designed and built by a number of groups.
\end{abstract}

\section{Introduction}
Detection of terrestrial-mass extra-solar planets (exoplanets) requires extremely stable and precise instruments. In the strive to reach extreme levels of Doppler precision, a number of previously uncharacterized instrumental effects will begin to dominate instrument error budgets. Dedicated Doppler radial velocity (RV) instruments must be able to measure shifts in stellar spectra at precisions approaching 10 c\ms in order to detect true Earth twins orbiting nearby stars. In the pursuit of reaching this bold precision goal, a number of previously uncharacterized instrument noise sources must be studied and mitigated at unprecedented levels. In the past decade, much attention has been paid to maximizing detection sensitivity by using optical fibers to deliver light from the telescope to the spectrograph. Typical multi-mode fibers (MMFs) used in current generation \lq{}seeing-limited\rq{} Doppler instruments have the convenient ability to \lq{}scramble\rq{} light, producing an output illumination that is significantly, though not perfectly, decoupled from the incident flux distribution \citep{Sturmer:2014, Halverson:2015}. However, these fibers do suffer from mode interference effects, commonly referred to as \lq{}modal-noise\rq{}, which can impose a fundamental limit on achievable measurement precision if not addressed properly \citep{Baudrand:2001}. This has been shown to severely limit both achievable signal-to-noise on stellar targets \citep{Iuzzolino:2014}, and ability to realize the full potential of coherent frequency calibration sources \citep{Ycas:2012}. It is important to note that insufficient image scrambling and fiber modal noise can both limit measurement precision, but are fundamentally different phenomena, as described in \cite{Halverson:2015}. As precision goals approach 10 cm s$^{-1}$, a number of subtle and largely unexplored instrumental systematics will begin to dominate overall performance error budgets. 

More recently, the use of SMFs for compact Doppler spectrometers has been suggested as a method of minimizing instrument size while overcoming many of these systematics  \citep{Schwab:2012, Crepp:2014b}. These fibers support only a single spatial propagation mode and therefore do not suffer from the classical modal interference effects of MMFs. The output intensity distribution of a SMF is entirely decoupled from input illumination variations, yielding a fundamentally stable instrument point-spread function (PSF) with perfect radial and azimuthal scrambling. These qualities make SMFs an attractive choice when designing compact, stabilized Doppler velocimeters for instruments on small telescopes \citep{Feger:2014, Blake:2015}, where mode-matching is relatively efficient, and for larger telescopes with high performance adaptive optics (AO) systems working near the diffraction limit \citep{Schwab:2012, Jovanovic:2014}. 

\begin{figure*}
\begin{center}
\includegraphics[width=7in]{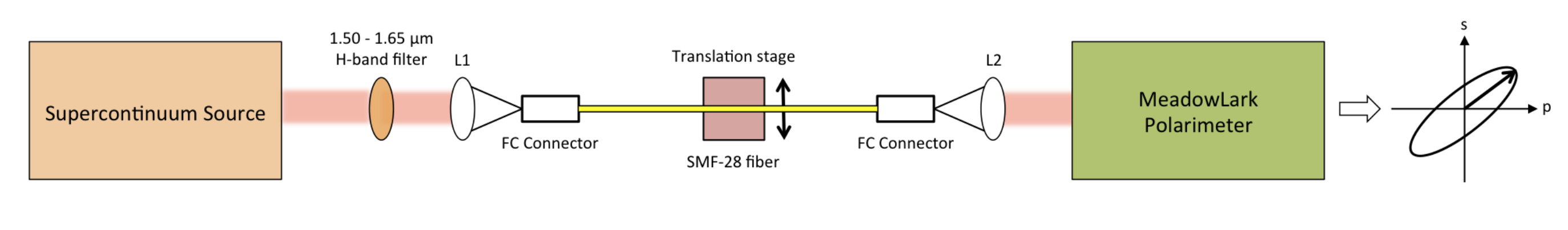}
\caption{Laboratory measurement setup. Light from a weakly ($\sim$10\%) polarized supercontinuum source is coupled into an SMF-28 patch cable. The polarization exiting the fiber is measured with a MeadowLark PMI-NIR polarimeter as the fiber is physically perturbed.}
\label{fig:setup}
\end{center}
\end{figure*}

While these fibers can produce a stable and perfectly scrambled output illumination, typical SMFs do support two fundamental polarization modes. Similarly to MMFs, imperfections and internal stresses in the fiber can lead to variable coupling between these two polarization modes \citep{Monerie:1980fk}. In SMFs, this leads to an effective polarization change of propagating light. In fact, a single fiber cable can very efficiently rotate or circularize any arbitrary input polarization to an arbitrary output polarization. As such, stressed single-mode fiber cables are frequently used as in-line wave-plates for polarization rotation of monochromatic sources \citep{Koehler:85}.

\begin{figure}
\begin{center}
\includegraphics[width=3.4in]{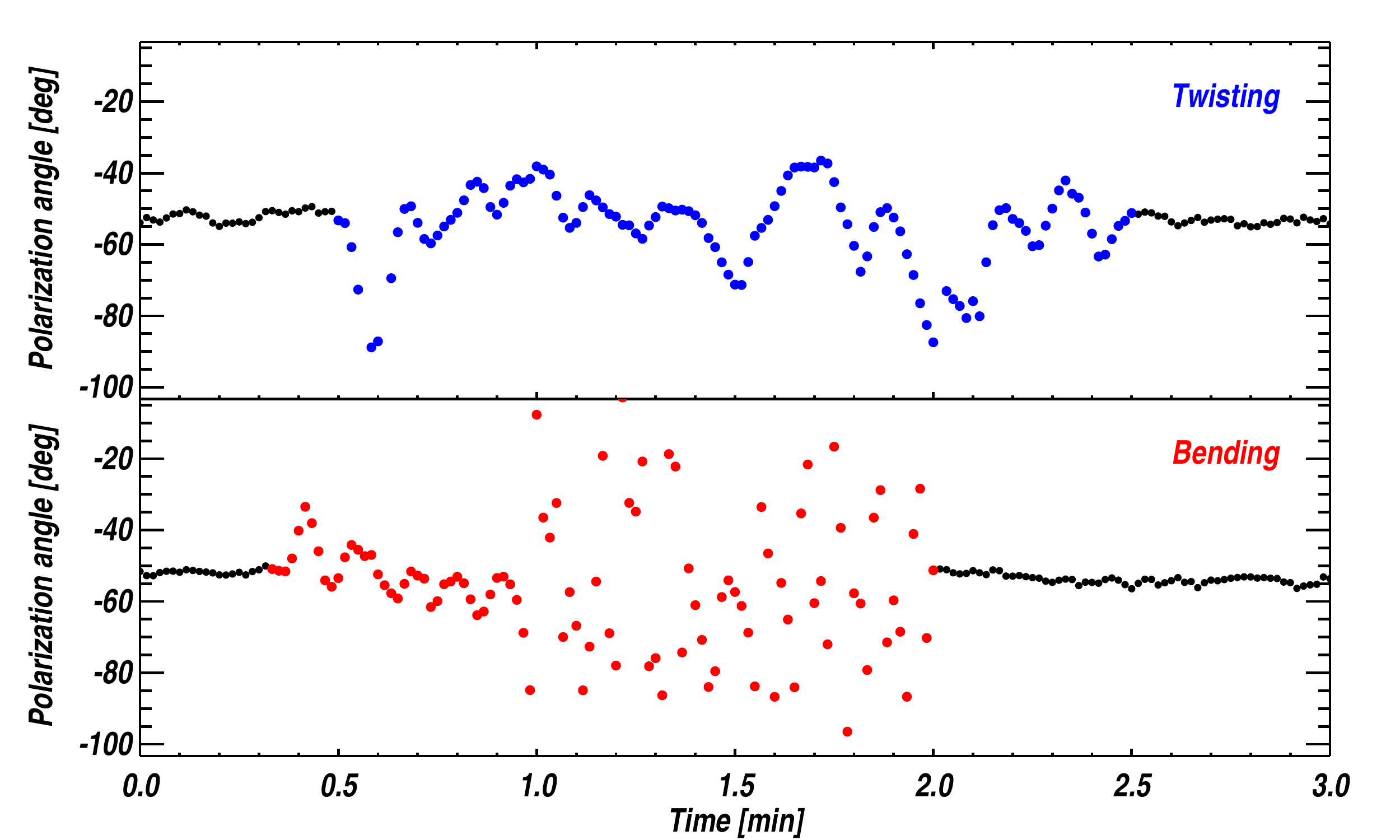}
\caption{Laboratory measurements of stress-dependent polarization mode coupling in a 2 meter SMF-28 fiber illuminated by a weakly ($\sim$10\%) polarized supercontinuum source. Top: Effective polarization rotation induced by variable twisting of the fiber cable. Bottom: Bend-induced polarization rotation measurement. In both cases, the effective polarization angle exiting the fiber varied by 10's of degrees with minimal stress applied.}
\label{fig:lab_data}
\end{center}
\end{figure}

Here we study the impact of polarization rotation in single-mode fibers on radial velocity measurements by examining the polarization sensitivity of the primary disperser (in this case, an echelle reflection grating) used in most Doppler spectrometers. The purpose of this study is to emphasize the scope and magnitude of this effect, and show specifically how this affects high precision RV measurements. This is of particular importance since there are several instruments currently being designed or built that utilize SMF delivery systems, including MINERVA-red \citep{Blake:2015}, iLocator \citep{Crepp:2014}, and tests of an extreme precision Doppler spectrometer for the Subaru telescope \citep{Schwab:2012, Jovanovic:2014}.

\section{\lq{}Modal-noise\rq{} in single-mode fibers}
With any fiber-fed instrument, the internal stresses within the fiber will change as the fiber is perturbed (e.g. due to telescope tracking, temperature variations, etc.) This variable stress can change the intrinsic fiber birefringence, which alters the polarization distribution exiting the fiber. The consequences of this variable birefringence have been studied for interferometric applications (e.g. \cite{Anderson:2014}), as birefringent effects in standard SMFs can degrade fringe contrast \citep{Kotani:2003}, but they have yet to be thoroughly studied in the context of precision Doppler spectroscopy.

The goal of this study is to estimate how these birefringence effects propagate to spectroscopic velocity errors. As such, we do not present a rigorous mathematical model of fiber birefringence in typical SMFs, as this has been abundantly documented and studied in the literature over the past several decades (e.g. \cite{Chartier:01,Rashleigh:1983}). Instead, we present a brief summary of the scope of the problem and the relevance to astronomical spectroscopy. 

In SMFs, any stresses induced on the fiber core, e.g. due to variations in applied pressure on the cable \citep{1071213}, bends \citep{Ulrich:80}, twists \citep{Ulrich:79, Smith:80}, thermal fluctuations \citep{Smith:80}, or even variations in external magnetic fields \citep{Rashleigh:81}, will lead to variable polarization mode coupling in the fiber core. This variable coupling will, in effect, rotate the polarization of light propagating through the fiber. Since most SMFs typically support two polarization modes, this variable mode coupling can lead to a manifestation of fiber modal noise. For astronomical instruments utilizing dedicated calibration fibers, any variations in the spectral profile due to variable polarization will not be traced with a calibration source, as the polarization state is specific to the source and fiber pair.  
It is important to note that any polarization rotation will also have a wavelength dependence, which depends strongly on the polarization \lq{}beat length\rq{} of the fiber in question \citep{Filippov:1990} and the internal and external stresses applied to the fiber core \citep{Eickhoff:81}.

Polarization-maintaining (PM) fibers have the ability to maintain a single polarization with a high polarization extinction ratio, but rely on stable and precise alignment of the incident polarization with the fast axis of the fiber for efficient mode propagation. This generally requires the incident illumination to be strongly linearly polarized, which is usually an impractical choice for most astronomical applications due to the associated efficiency losses.

\begin{figure*}
\begin{center}
\includegraphics[width=3.33in]{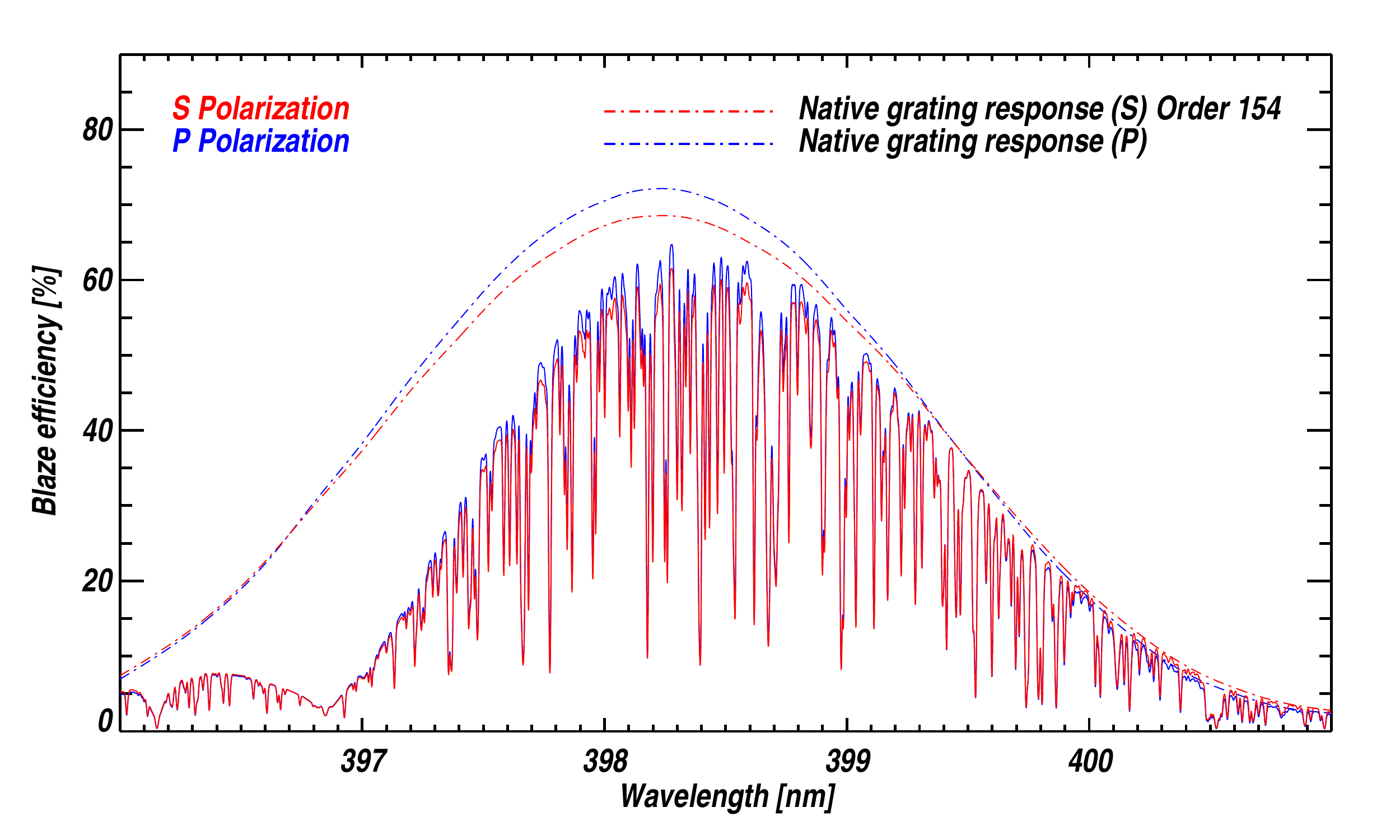}\includegraphics[width=3.33in]{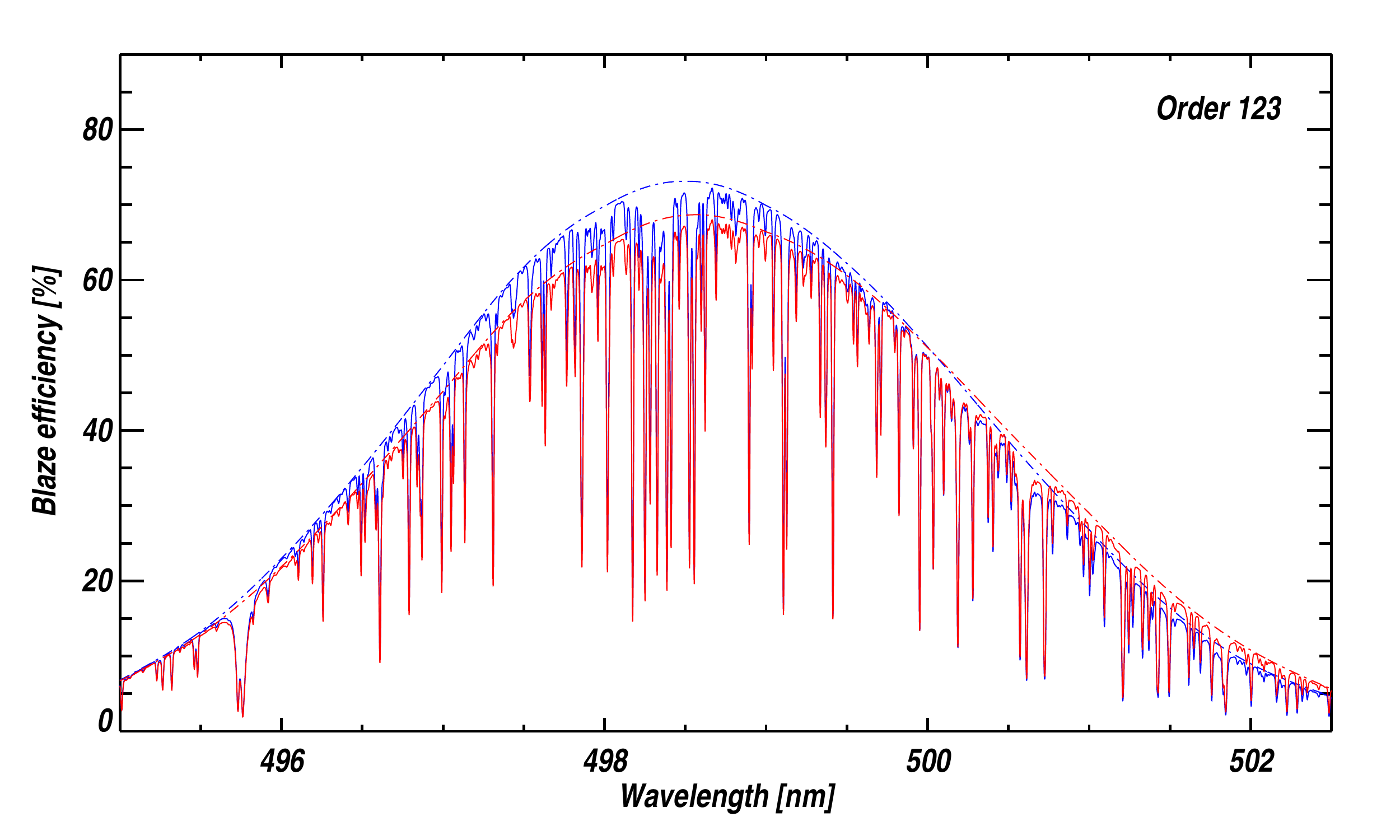}
\includegraphics[width=3.33in]{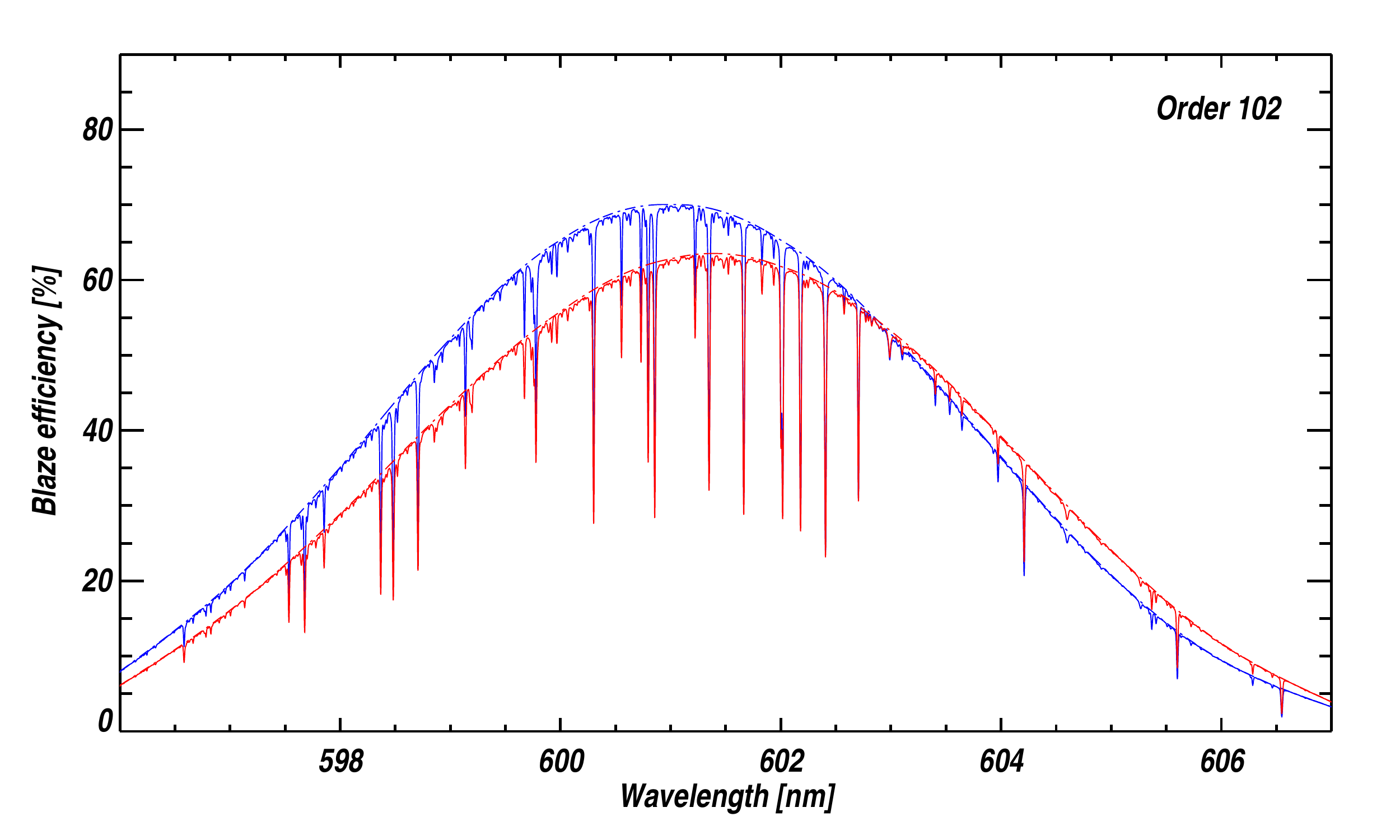}\includegraphics[width=3.33in]{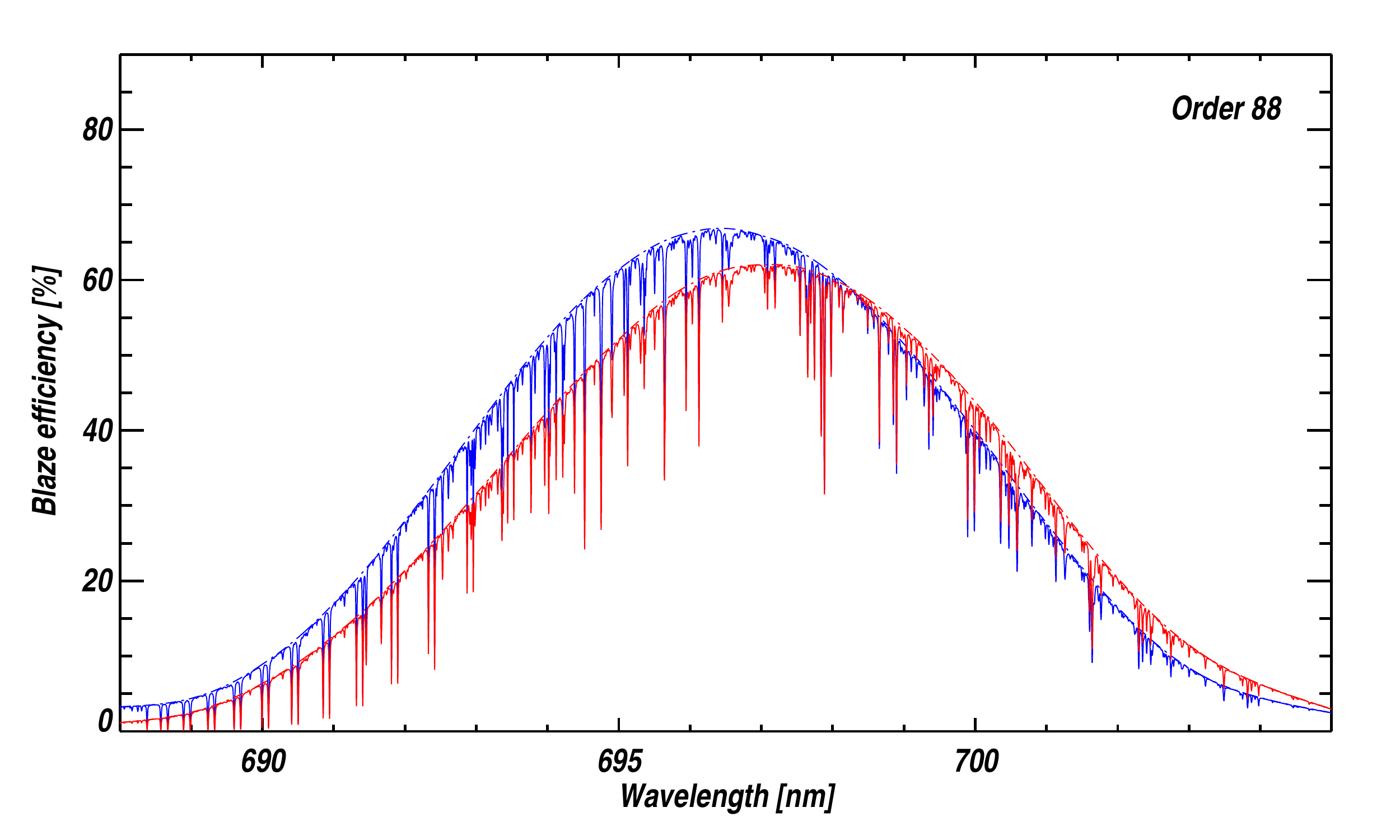}
\caption{Blaze efficiency functions for a variety of echelle orders of a standard Aluminum-coated Newport R4 echelle grating (MR160) multiplied by high resolution synthetic G2 spectrum. Blaze efficiency data provided by Newport RGL. Efficiency response functions for orthogonal polarization states are shown in red and blue. Model stellar spectra assume $R$=100,000 instrument resolution and 3-pixel sampling of the instrument PSF. Discrepancies between blaze efficiency curves and synthetic spectra are due to continuum suppression in certain spectral regions. Velocity offsets were measured by modulating the stellar spectrum with a linear combination of both $S$ and $P$ blaze curves. }
\label{fig:grating_response_vis}
\end{center}
\end{figure*}

\begin{figure*}
\begin{center}
\includegraphics[width=3.33in]{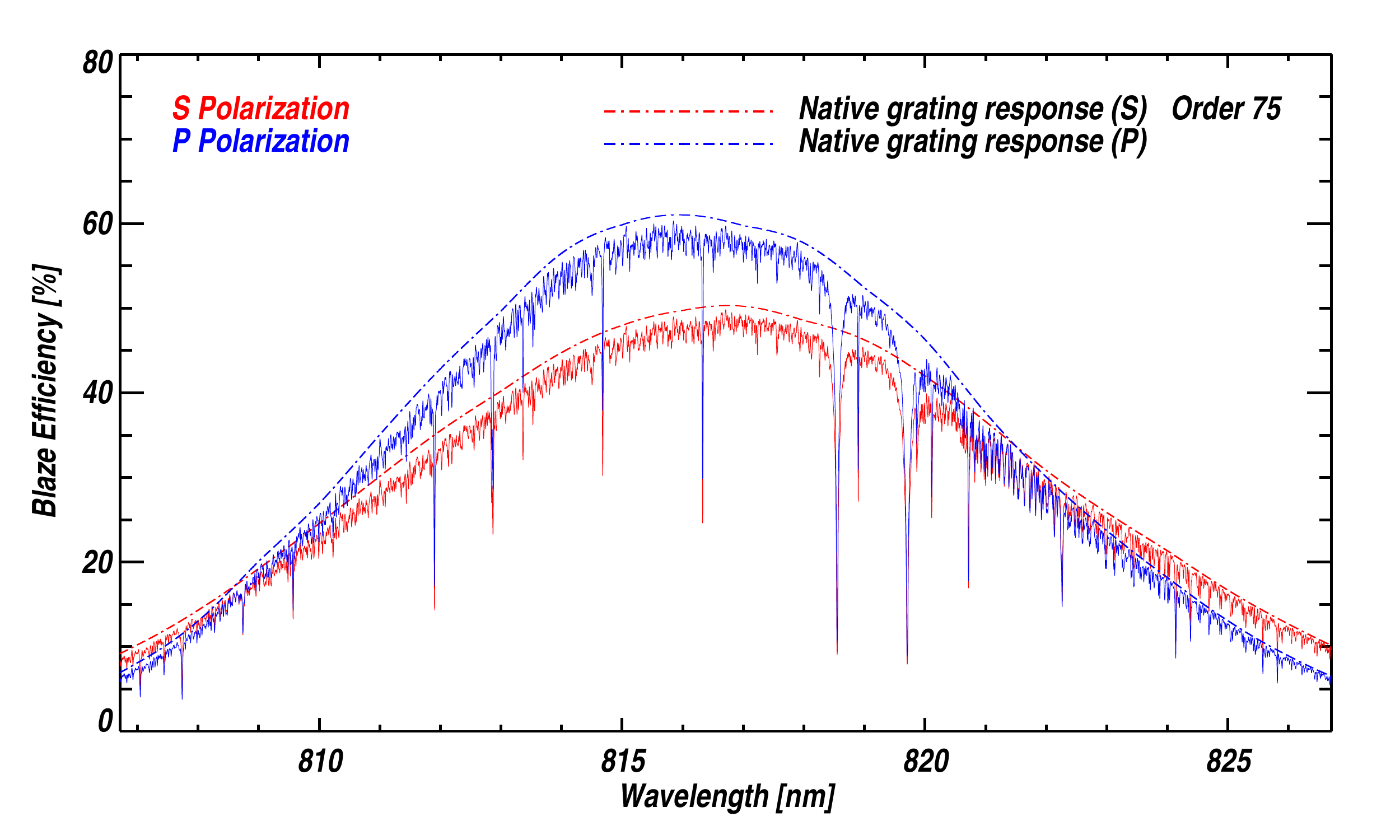}\includegraphics[width=3.33in]{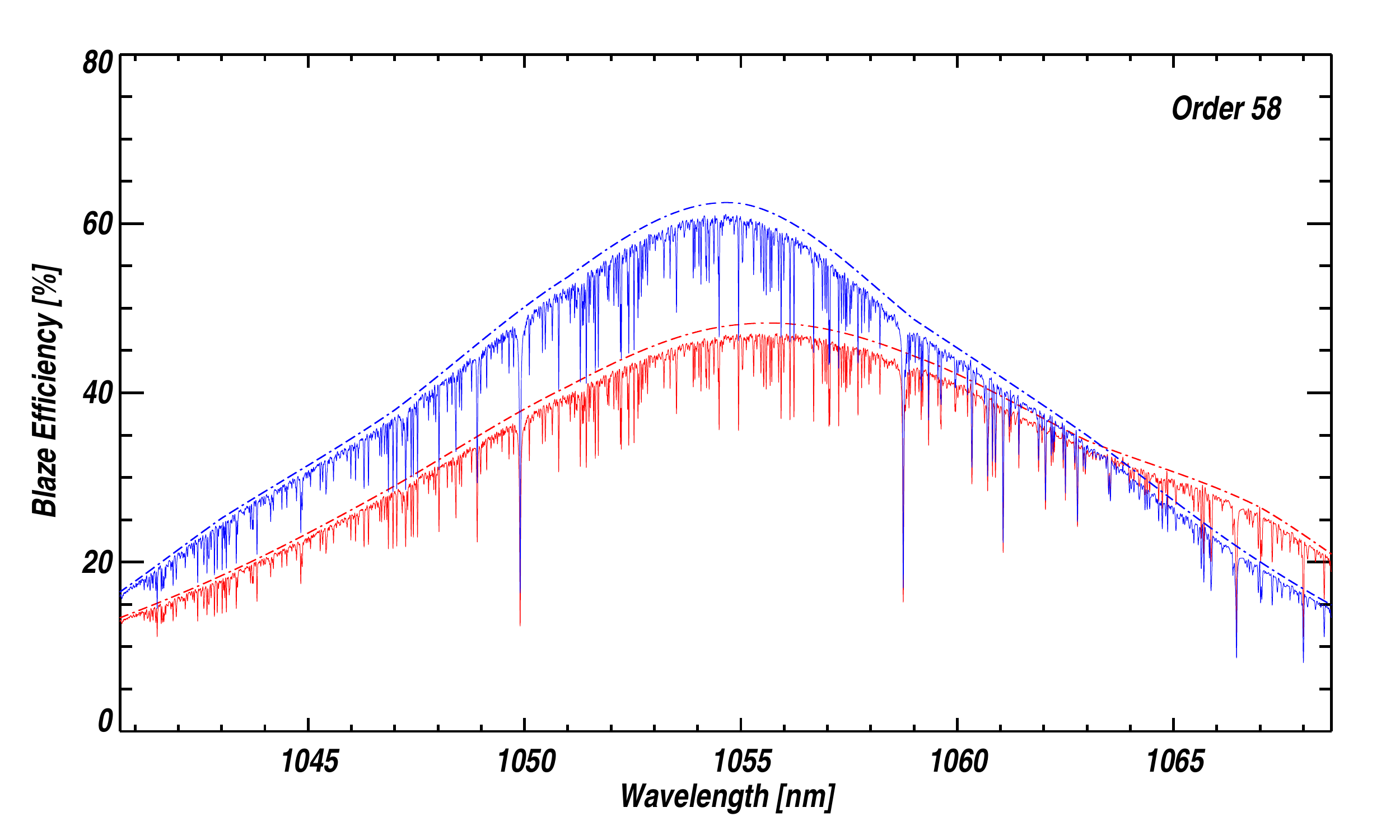}
\includegraphics[width=3.33in]{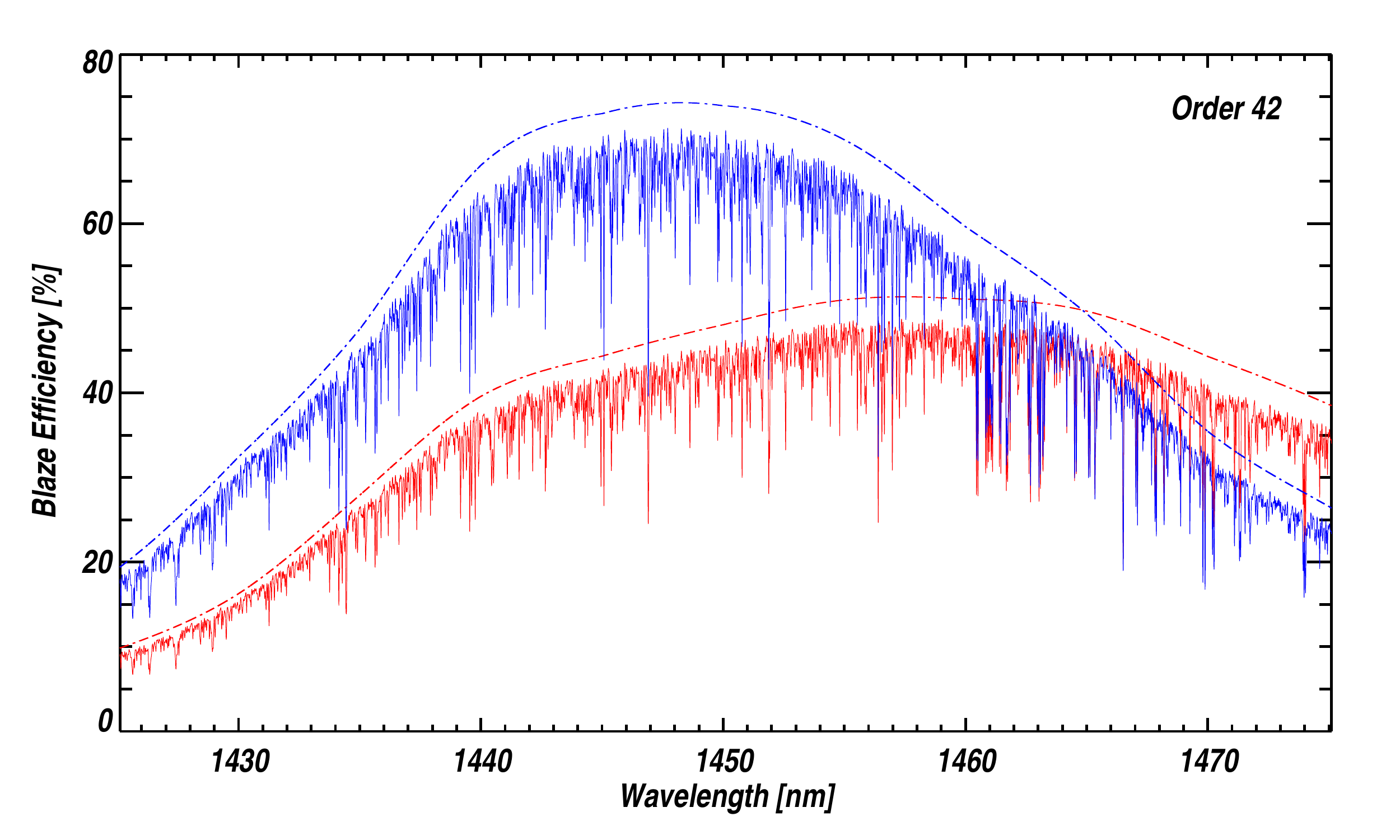}\includegraphics[width=3.33in]{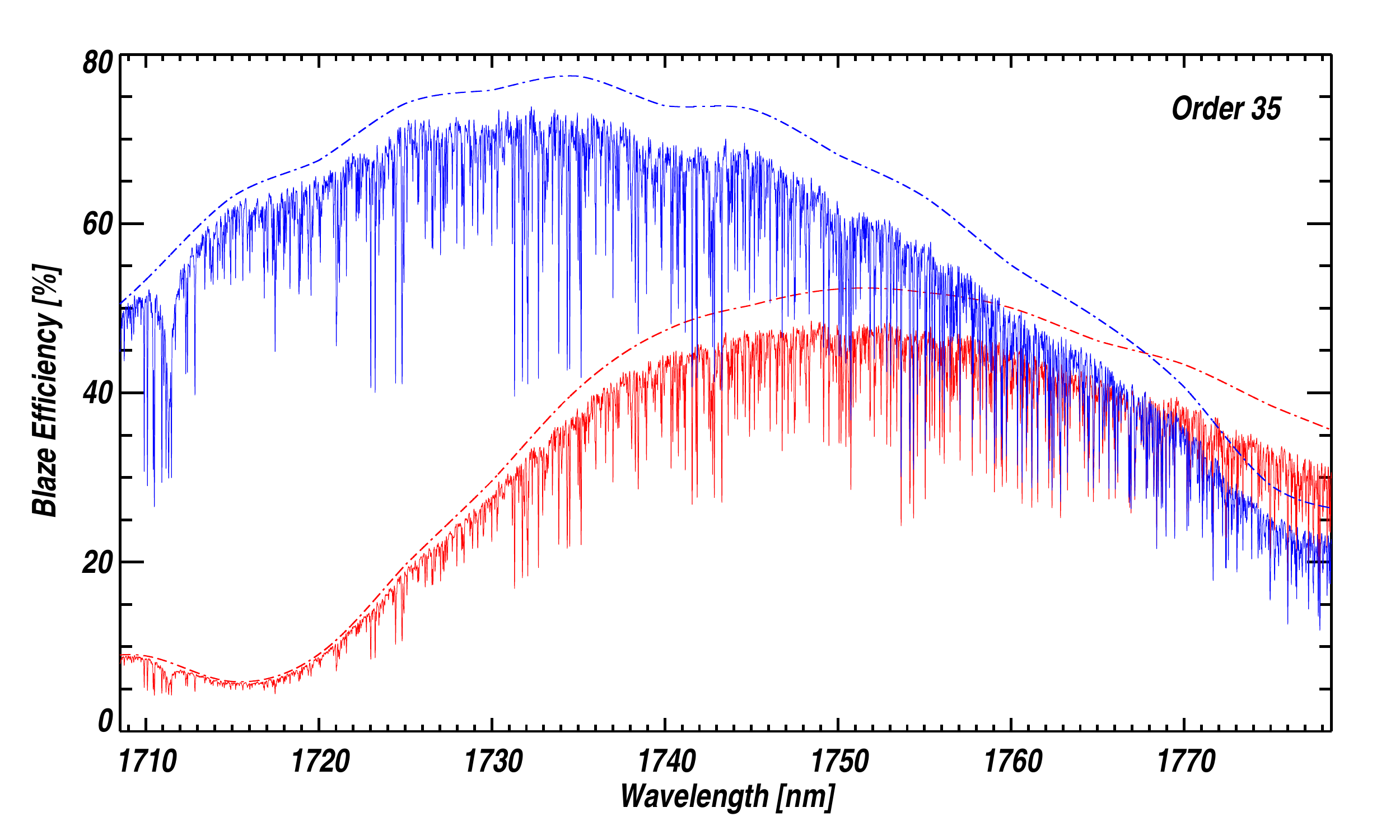}
\caption{Similar to previous plot, though for a different Newport R4 echelle grating (MR257). Order blaze efficiencies are multiplied by a synthetic 3000 K, M Dwarf spectrum. Model spectra parameters are identical to the previous case.}
\label{fig:grating_response_nir}
\end{center}
\end{figure*}

\subsection{Laboratory measurements}
While ample amounts of literature exists that studies the effects of stress-induced polarization rotation in SMFs (see previous references), we perform a series of simple laboratory tests to show the magnitude of possible polarization fluctuations expected from typical stresses associated with astronomical telescope fiber feeds (namely twists and bends effects) illuminated by a weakly polarized continuum source, meant to emulate a possible astrophysical target.

A 2 meter commercial SMF cable (Corning SMF-28) is illuminated by a supercontinuum source that is intrinsically weakly polarized (degree of polarization is approximately 10\%, as measured by our polarimeter). The output polarization of the fiber cable is then recorded using a commercial PMI-NIR MeadowLark Polarimeter. An H-band interference filter (1.5 - 1.65 $\mu$m) is inserted at the fiber exit to ensure the illumination bandpass is restricted to wavelengths where the fiber supports a single spatial mode (1.26 - 1.65 $\mu$m). The fiber is coiled onto a linear translation stage, which allows for a variable stress to be induced in the cable, and terminated with standard telecommunication FC connectors on either side.  A schematic of the experimental setup is shown in Figure~\ref{fig:setup}. The cable is then perturbed by both translating the linear stage (inducing a mild bend) and twisting the fiber by hand. For twist measurements, the cable was periodically twisted by hand between 0 and 90 degrees across a 0.5 meter section of the cable. For bend measurements, the cable was bent between two fixed mounting points, separated roughly 0.3 meters apart, using a mechanical stage. This varied the overall fiber bend radius by several inches. 

We find that, even for low-levels of added stress, the effective output polarization angle of light exiting the fiber is highly variable. Measurements of the fiber output polarization are shown in Figure~\ref{fig:lab_data}.  Mild twists and bends of the fiber cable can induce effective polarization rotations of over 10 degrees. 

\section{Impact on measured radial velocities}

To estimate the direct effects of polarization variations on measured radial velocities, we focus on the polarization sensitivity of the primary dispersive element in the spectrograph: in this case, an R4 echelle grating. We use measured $S$ and $P$ polarization curves for two Newport RGL gratings, manufactured with two different grating masters: MR 160, used on HARPS \citep{Mayor:2003}, and MR 257, used on HPF \citep{Mahadevan:2014} and CARMENES \citep{Quirrenbach:2014} instruments, to modulate simulated stellar spectra (see Figures~\ref{fig:grating_response_vis} \& \ref{fig:grating_response_nir}). The modulation is a linear combination of different ratios of the measured grating blaze functions for the two orthogonal incidence polarizations, for a variety of spectral orders in the visible and near-infrared (NIR).

Model spectra are assumed to be noiseless, and are generated at an effective instrument resolution of $R$=100,000. For the MR160 grating, we use a synthetic G2 spectrum (T$_{\mathrm{eff}}$ = 5700 K) for radial velocity calculations for four spectral orders between 350 and 700 nm. For the MR257 grating, where our available data is centered in the NIR, we use a synthetic M-dwarf spectrum as a stellar template (in this case a 3000 K BT-Settl model).

The variable blaze modulation due to changing polarization can induce apparent velocity offsets. On the smallest scale, it can introduce a variable gradient across individual spectral features, although this is likely a negligible change. Across an order, it changes the relative chromatic weighting of lines, and shifts the center of the cross-correlation peak in that wavelength range. Most importantly, it changes the shape of the aggregate cross-correlation function, and the final determination of radial velocity from a single observation. These effects are similar to what would be observed due to variable atmospheric refraction effects (e.g. due to airmass changes, variations in atmospheric dispersion correction, etc.).

\begin{figure*}
\begin{center}
\includegraphics[width=3.4in]{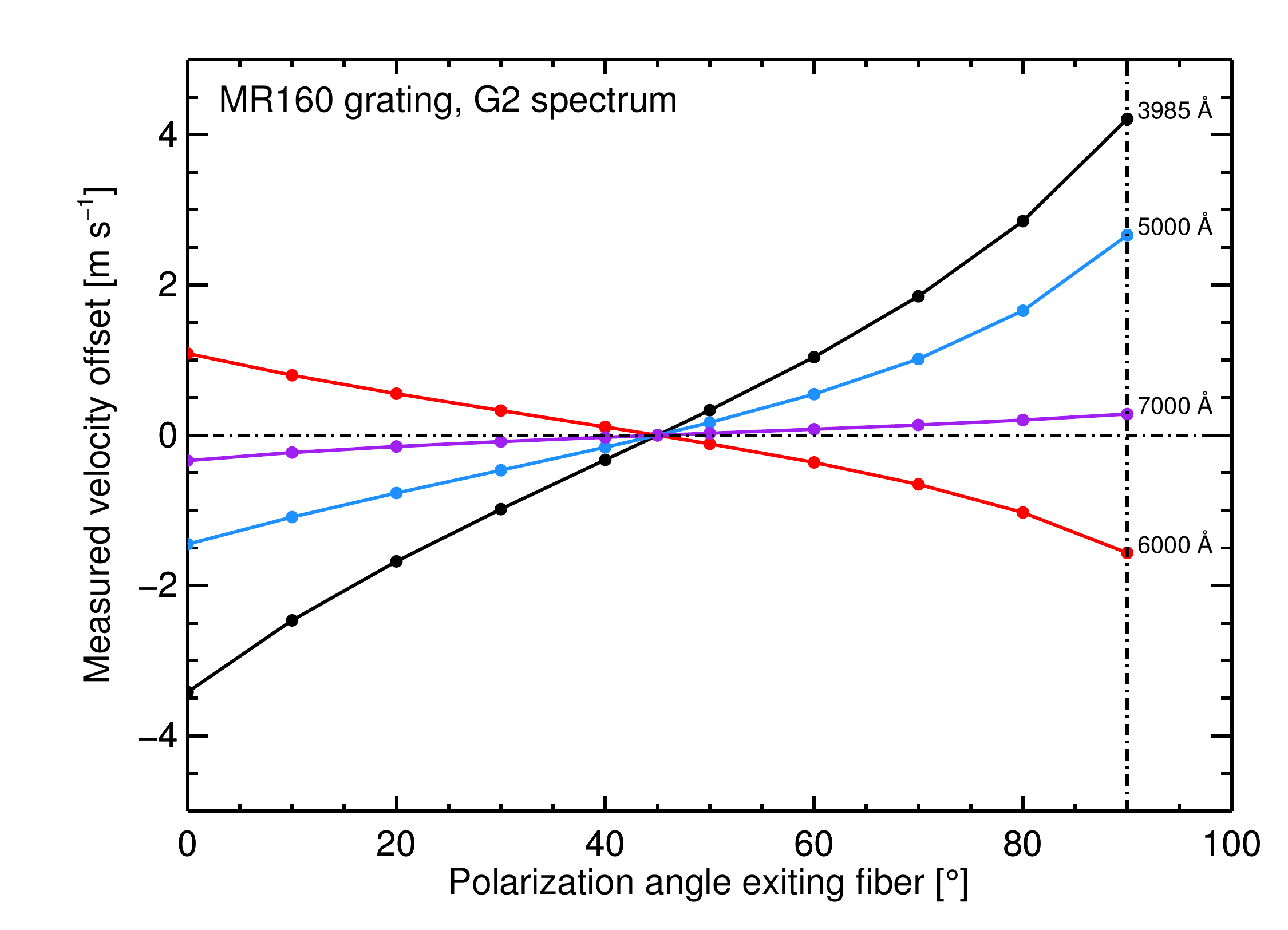}\includegraphics[width=3.4in]{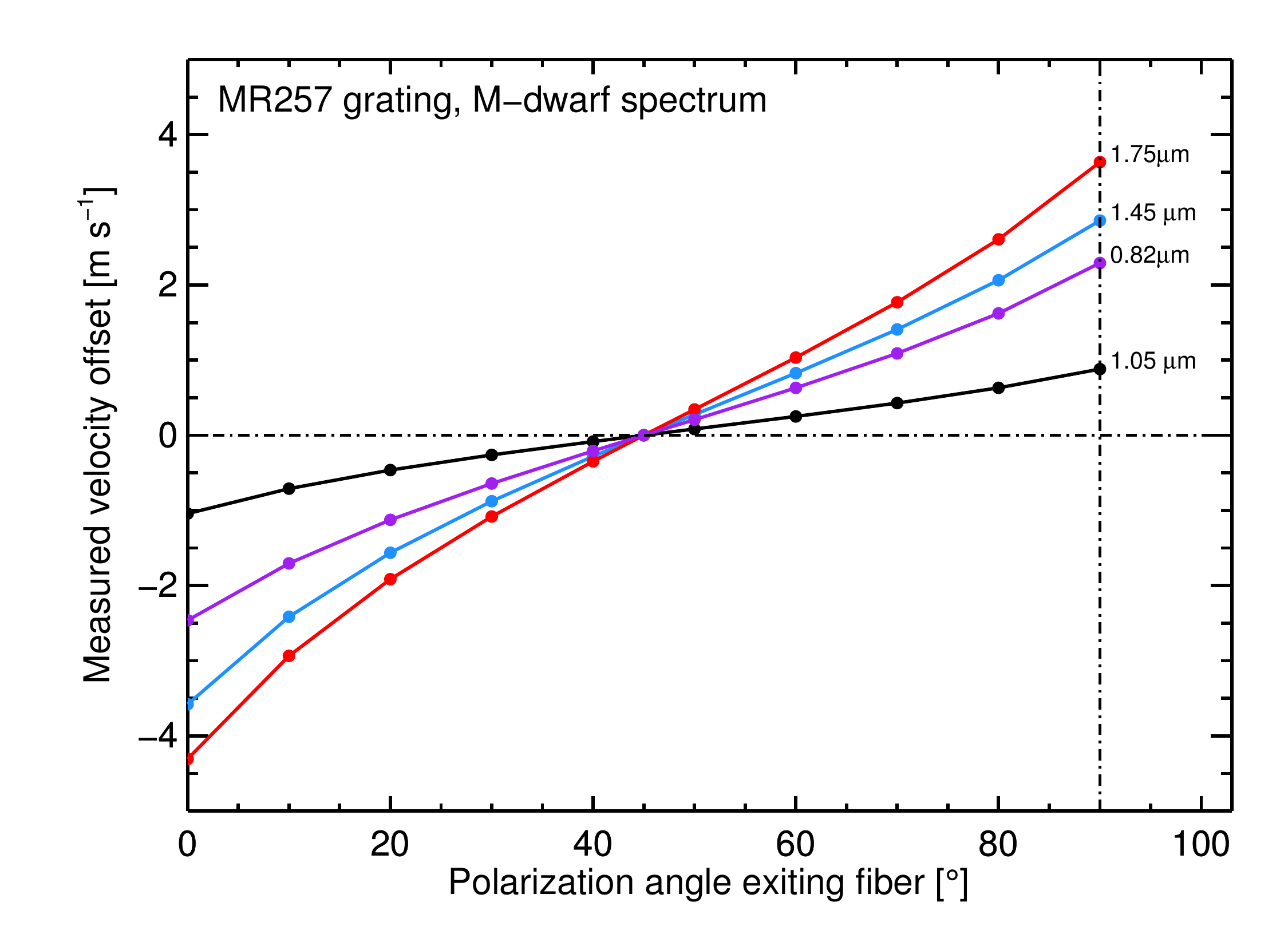}
\caption{Calculated velocity offsets for synthetic spectra modulated by polarization-dependent echelle efficiency curves. Wavelengths for each of the four orders are listed on the right. All velocity offsets are calculated relative to a 45 degree effective polarization incident on the echelle. Left: Velocity offsets for the MR160 grating using a synthetic G2 stellar spectrum. Right: Offsets for the MR257 using a synthetic M-dwarf spectrum.}
\label{fig:rv_offset}
\end{center}
\end{figure*}

Relative velocity offsets are calculated using weighted mask cross-correlation method (similar to the methods described in \cite{Baranne:1996} and \cite{Chakraborty:2014}) with a custom stellar template mask built specifically for this purpose for each spectral class considered (approximately G2 and M5). For the latter, telluric contamination is neglected since a synthetic spectrum is used, but this could be easily adapted for use with an observed spectrum with a slightly more rigid selection procedure for mask lines. Velocity offsets at all effective polarization combinations are then compared to spectra modulated by an \lq{}unpolarized\rq{} blaze function (average of $S$ \& $P$). Spectra for each order were purposely not continuum normalized when calculating velocity offsets, as the blaze functions contain important information relating to the weights of various spectral features used for the cross-correlation. 

Measured velocity offsets between polarization angles are shown in Figure~\ref{fig:rv_offset}. We note that effective shifts of several \ms are seen between orthogonal polarizations for nearly all measured echelle orders. Even for weakly polarized light, low-level polarization rotation due to variable mode coupling can still introduce errors at the 10's of c\ms level. For strongly polarized, coherent sources, such as laser frequency combs, this effect will be even more pronounced. If not dealt with, this systematic error could conceivably compromise the exquisite stability capabilities of dedicated SMF Doppler velocimeters. 

\section{Possible Solutions}
\subsection{Mechanical agitation}
Periodic mechanical agitation of the fiber cable will change the polarization mode coupling of incident radiation. This method is frequently used to temporally vary the phase distribution of modes in MMFs, which can significantly reduce classical mode interference effects \citep{Mccoy:2012, Lemke:2011}. In SMFs, this agitation could conceivably \lq{}scramble\rq{} the effective mode coupling. This could reduce the effective degree of polarization over typical astronomical integration times, though may result in an efficiency drop due to increased bend losses depending on the amplitude and frequency of agitation.

It has been shown that using long fiber cables can effectively act as \lq{}depolarizers\rq{} for strongly polarized sources \citep{Malykin:2009uq,1072087}, though likely not at the level required to completely mitigate this issue for precise RV measurements. 

\subsection{Software corrections}
In addition, it may be possible to mitigate this effect post factum through continuum normalization or ``deblazing'' of the spectrum before calculating RVs. For example, using the continuum normalization routines included in the {\it REDUCE} package \citep{Piskunov:2002}, we are able to mitigate the RV disparity by about an order of magnitude. Depending on the precision level desired, this might be the most convenient solution, but will remain sensitive to the normalization procedure, and might produce RV offsets as the fits to the varying order response functions change. Additionally, important spectral weighting information, such as absolute chromatic flux level across a given spectral order, is omitted by dividing by a estimated blaze function.

\section{Applicability to multi-mode fiber fed instruments}
While we have focused entirely on standard SMFs in this study, it is import to note that the polarization effects discussed here could conceivably affect multi-mode fiber delivery systems at the c\ms level. This is particularly true for instruments using small diameter, asymmetric core geometry fibers. These fibers can significantly improve scrambling performance over standard circular-core fibers, but may also be prone to maintaining measurable levels of residual polarization \citep{Avila:2014}\footnote{\url{http://web.mit.edu/gfuresz/www/FOiA_IV/Fiber_Optics_in_Astronomy_IV./Presentations.html}}. 

HARPS has observed subtle, though measurable, variations in measured blaze functions at the $\sim$10$^{-3}$ level \citep{Piskunov:2010}\footnote{\url{http://exoplanets.astro.psu.edu/workshop/presentation/4-b-Piskunov-DataReduc.pdf}}. While these variations could be attributed to a number of error sources (such as seeing variations, telescope focus drifts, airmass variations, etc.), residual polarization in the instrument delivery fiber could also be a contributing factor, and should be considered along with more commonly examined sources of variability.

\section{Summary}

We present a cautionary note that variable coupling of polarization modes in single-mode fibers can cause systematic shifts in measured radial velocities for spectrographs using standard echelle reflection gratings as the primary disperser. This effect is comparable to classical modal interference in multi-mode fibers, though it manifests in a more subtle manner. We perform a simple laboratory test to explore the effects of external stresses on polarization mode coupling by measuring the output polarization of a commercial SMF illuminated by a weakly polarized continuum source under external stress. To quantify the magnitude of this effect on RV measurements, we calculate the effective Doppler shift of a template spectrum modulated by measured echelle grating response functions for several spectral orders for two different manufactured gratings. We use a linear combination of these response functions to span a wide range of effective polarizations incident on the grating. We show that, even for low-level rotation of weakly polarized light, RV offsets of 10's c\ms are entirely possible for instruments using SMFs. This subtle effect is not traced with a simultaneous calibration source, and can therefore lead to insidious systematic shifts in stellar radial velocity measurements. In the era of extreme precision Doppler spectrometers striving towards 10 c\ms precision, this effect could conceivably dominate measurement error for SMF-based instruments.

\acknowledgments{This work was partially supported by funding from the Center for Exoplanets and Habitable Worlds. The Center for Exoplanets and Habitable Worlds is supported by the Pennsylvania State University, the Eberly College of Science, and the Pennsylvania Space Grant Consortium.  We acknowledge support from NSF grants AST 1006676, AST 1126413, AST 1310885, and the NASA Astrobiology Institute (NNA09DA76A) in our pursuit of precision radial velocities in the NIR. SPH acknowledges support from the Penn State Bunton-Waller, and Braddock/Roberts fellowship programs and the Sigma Xi Grant-in-Aid program. We thank Richardson Grating Labs for providing blaze efficiency curves for a sample R4 echelle grating for a variety of spectral orders.}
\newpage


\end{document}